\documentclass[
 aip,
 amsmath,amssymb, 
 reprint,longbibliography
]{revtex4-1}
\usepackage{graphicx}
\usepackage{tabularx}

\begin{document}

\title{Implementation and performance of a fiber-coupled CMOS camera in an ultrafast reflective high-energy electron diffraction experiment} 

\author{\firstname{Jonas D.} Fortmann}
\affiliation{Faculty of Physics, University of Duisburg-Essen, 47057 Duisburg, Germany}

\author{Alexander Kaßen}
\affiliation{Faculty of Physics, University of Duisburg-Essen, 47057 Duisburg, Germany}

\author{Christian Brand}
\affiliation{Faculty of Physics, University of Duisburg-Essen, 47057 Duisburg, Germany}

\author{Thomas Duden}
\affiliation{Th. Duden Konstruktionsbüro, Mustangweg 17, 33649 Bielefeld, Germany}

\author{Michael \surname{Horn-von Hoegen}}
\email[Corresponding author: ]{horn-von-hoegen@uni-due.de}
\affiliation{Faculty of Physics, University of Duisburg-Essen, 47057 Duisburg, Germany}
\affiliation{Center for Nanointegration (CENIDE), University of Duisburg-Essen, 47057 Duisburg, Germany}

\date{\today}

\begin{abstract}

The implementation of a monolithic fiber-optically coupled CMOS-based TemCam-XF416 camera into our ultra-high vacuum (UHV) ultrafast reflection high-energy electron diffraction setup is reported. 
A combination of a pumpable gate valve and a self-built cooling collar allows UHV conditions to be reached without the need to remove the heat-sensitive device.
The water-cooled collar is mounted to the camera housing and prevents heating of the detector upon bake-out of the UHV chamber.
The TemCam provides an one order of magnitude higher spatial resolution than the previously used microchannel plate (MCP) based detector (Burle Chevron 3040FM) which enables a 30\,\% higher resolution in reciprocal space.
The low background intensity and the 4$\times$  lager dynamic range enables analysis of the diffuse intensity of the diffraction pattern like Kikuchi lines and bands.
A key advantage over the previous MCP detector is the complete absence of the blooming effect, which enables the quantitative spot profile analysis of the diffraction spots. 
The inherent light sensitivity in an optical pump experiment can be overcome by using photons with $h\nu < 1.12$~eV, i.e., the indirect band gap of silicon, or by shielding any stray light.
\end{abstract}

\maketitle

\section{Introduction}

With the commercial availability of modern imaging electron detectors with high pixel resolution for application in transmission electron microscopes, there is growing interest in using these cameras for experiments under ultra-high vacuum (UHV) conditions \cite{niuMAXPEEMSpectromicroscopyBeamline2023,janoschkaImplementationOperationFibercoupled2021,waldeckerTimedomainSeparationOptical2015,grobRankingTEMCameras2013, tysonSurfaceMagnetismFe3GeTe22024,ruskinQuantitativeCharacterizationElectron2013,rauchNewFeaturesCrystal2021,henkeIntegratedPhotonicsEnables2021,takInjectionMechanismsIIInitride2023,mattesFemtosecondElectronBeam2024,faruqiElectronicDetectorsElectron2011,weberElectronImagingNanoscale2024,hugenschmidtDirectSynthesisZIF82021,minenkovAdvancedPreparationPlanview2022}
These monolithic fiber-optically coupled CMOS-based detectors are fully integrated with a thin polycrystalline phosphor scintillator as the electron-sensitive component. High-energy electrons excite a shower of secondary electrons in the Al cover layer, which subsequently generate so many photons in the scintillator that individual events on the CMOS chip can be assigned to individual electrons \cite{stumpfDesignCharacterization162010}. 

Here, we report on the upgrade of our  ultrafast reflection high-energy electron diffraction experiment (URHEED) \cite{janzenUltrafastElectronDiffraction2006,
hanisch-blicharskiUltrafastElectronDiffraction2013,
frigge_optically_2017,
hafke_pulsed_2019,
hanisch-blicharski_violation_2021,horn-vonhoegenStructuralDynamicsSurfaces2024}
from a microchannel plate (MCP) detector (Burle Chevron 3040FM) to a fiber-optically coupled CMOS-based detector (TVIPS TemCam-XF416) \cite{tietzAdvantagesFiberOpticallyCMOS2022}.

The detective quantum efficiency of the TemCam is $\rm{DQE}(0)= 0.62$ \cite{shiCollectionMicroEDData2016} with a filling factor of 0.72 \cite{stumpfDesignCharacterization162010}. 
In contrast, the DQE of the MCP is somewhat smaller, as the channels cover only 55\% of the MCP surface (filling factor of 0.55) and the probability to generate an electron cascade by multiplication in general decreases with electron energy \cite{burleindustriesincProceduresSpecicationsAPD2001,woodheadChannelElectronMultiplier1977,klingelhoeferMeasurementDetectionEfficiency1986}. 
Thanks to the improved DQE of the TemCam, this camera is better suited than the MCP for recording the few most precious electrons in an ultrafast diffraction experiment.

We present a procedure for integrating the TemCam into an existing UHV apparatus and characterize its performance in comparison to the previously used MCP detector.

\section{Detector integration and handling}

\begin{figure}[htbp]
\centering
\includegraphics[width=1.000\columnwidth]{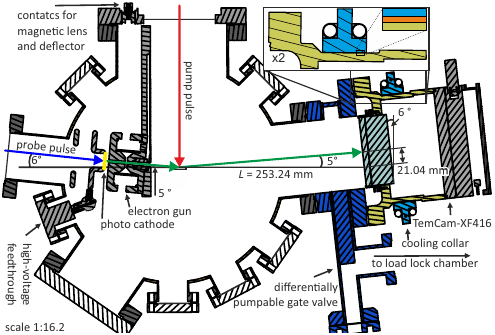}
\caption{
\textbf{URHEED setup} as cross section, with magnification of the cooling collar in the top right. Dark blue: differential pump-able gate valve, yellow: detector housing, light blue: cooling collar, gray: hard solder capillary, orange: copper tape.
}
\label{fig:URHEED_Setup}
\end{figure}

Whilst the TemCam is compatible with UHV conditions, the challenge is to achieve these, as such detectors must not get hotter than 40\textdegree C under any circumstances, otherwise serious damage can occur.
Establishing UHV conditions, however, requires prolonged bake-out of the entire chamber to temperatures well above 100\textdegree C which is incompatible with the maximum temperature requirements of the camera.  
A possible solution may be to dismount the detector before bake-out but involves considerable drawbacks \cite{janoschkaImplementationOperationFibercoupled2021}.
The detector may be physically damaged during assembly and reassembly.
It is exposed to ambient conditions and moisture that enter the UHV chamber after reconnecting the detector. 
Upon dismounting and re-attaching the detector, the warming-up and cooling-down cycles introduce thermal stress to the fiber-CMOS unit that should be avoided.
With the help of a self-built water-cooled collar we cool the housing of the detector and thus permit bake-out of the entire chamber without dismounting and warming-up of the detector.  

In our setup shown in FIG.~\ref{fig:URHEED_Setup}, the detector is mounted to the main chamber (MC) via a gate valve (VAT CF-F 100 CF-F 160).
The gate valve exhibits an additional CF-F 40 flange on the detector side with an angle valve (VAT 28.4 CF-F 40) attached which is connected to a turbo molecular pump (Pfeiffer HiPace 80) at the load lock chamber (LL) with a base pressure better than $3\times10^{-8}$\,mbar.
This combination allows the detector to be pumped through the MC or connected to the LL without mutual interference.
During imaging operation of the detector the angle valve to the LL is closed.
The MC of the URHEED experiment is pumped by a turbo molecular pump (Pfeiffer TMU 521), an ion getter pump (Varian) and two getter pumps (SAES CapaciTorr\textregistered-D 400-2). 
The pressure is measured by means of an ion gauge (MKS Granville Phillips GP307).

In case of venting the MC the gate valve is closed, the detector is kept cold via its internal Peltier cooler, and is pumped by the LL. Exposure to ambient conditions is thus avoided.
While the gate valve is heated during bake-out to ensure that UHV conditions are reached, the detector is located outside the bake-out oven and is not heated by radiation or convection.
During bake-out, the gate valve is closed and thus protects the scintillator surface from any direct heat radiation. 

\begin{figure}[hbtp]
\centering
\includegraphics[width=1.000\columnwidth]{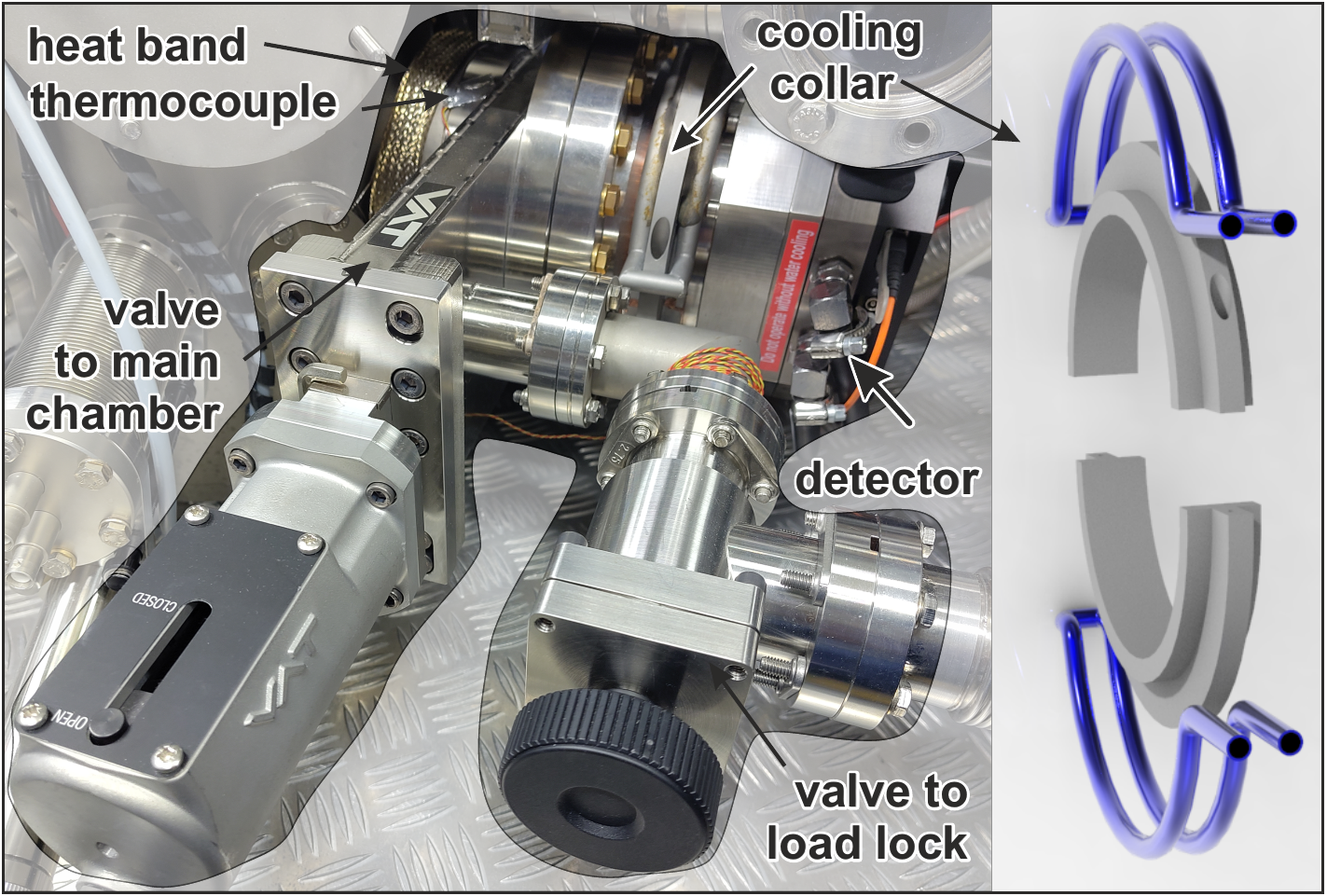}
\caption{
\textbf{Our setup} depicting the cooling collar (exploded view in the right panel), attached directly on the TVIPS TemCam-XF416 detector, which is mounted via a VAT gate valve CF-F 100 CF-F 160 at the main chamber. On the detector side an additional CF-F 40 pump flange is connected to our load lock system through an angle valve. On the side of the main chamber a heating tape is located and a type K thermocouple is installed in order to monitor the temperature.
}
\label{fig:Setup}
\end{figure}

To prevent heating of the detector through thermal conductivity via the hot gate valve, we mounted a water-cooled collar on the detector housing acting as heat sink.  
This collar is made of two parts of solid stainless steel half-rings, as shown in FIG.~\ref{fig:URHEED_Setup} and \ref{fig:Setup}.
Each half-ring is 30\,mm wide and 10\,mm thick. They are cooled by a capillary (stainless steel tube of 10\,mm outer and 9\,mm inner diameter), hard soldered for maximizing the thermal conductivity. 
Both capillaries are connected with PVC fiber-reinforced hose pipes in one single loop to the cooling water (15\textdegree C, 1.5~l/min).

\begin{figure}[htbp]
\centering
\includegraphics[width=1.000\columnwidth]{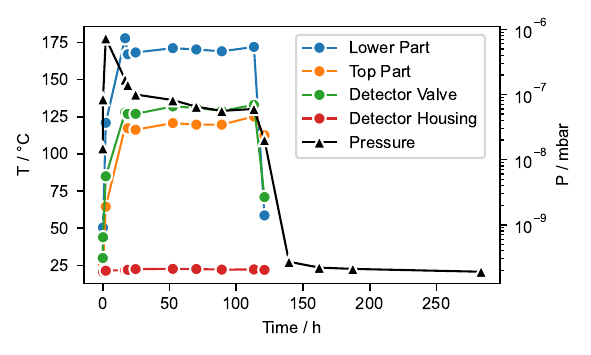} 
\caption{
\textbf{Temperature and pressure evolution during bake-out}. The temperature of the gate valve separating MC and detector was measured on the flange towards the MC. 
}
\label{fig:Baek_Data}
\end{figure}

The inside of each half-ring is covered with heat conducting copper tape to maximize the thermal contact across the 139.5\,$\rm{cm^2}$ contact area to the detector housing (illustrated in the magnification in FIG.~\ref{fig:URHEED_Setup}.
Both half-rings are tightly connected to each other through two M4 screws (tightened at 4\,Nm torque) and thus pressed onto the detector housing.
K-type thermocouples, which were attached at various points on the MC, were used to monitor the temperatures during the bake-out shown in FIG.~\ref{fig:Baek_Data}. 
The temperature of the detector housing (red data points and solid line) rises just one Kelvin from 21\textdegree C to 22\textdegree C while the temperature of the gate valve connected to the detector increases up to 130\textdegree C.
The low temperature rise by 1\,K proves the efficiency of our cooling collar.
The heating of the lower MC together with the ion getter pump and turbo molecular pump was turned off after 5 days, while the upper part of the MC followed 12 hours later.
Subsequently, the pressures dropped from $6\times10^{-8}$\,mbar to below $2\times10^{-10}$\,mbar (black triangles and line in FIG.~\ref{fig:Baek_Data}.
The detector was reconnected to the MC by opening the gate valve and closing the angle valve to the LL one day after bake-out.

In order to remove noise and detector artifacts -- originating from the coupling of the two fiber bundle at the vacuum junction and from the CMOS sensor -- from the images dark- and flatfielding is required.
While dark images can be recorded for all relevant exposure times by just averaging over images taken without any electron beam, flat images require a homogeneous illumination of the whole detector area in order to work properly.
In TEM or LEEM typically the electron source and the electron optics under strong defocussing conditions are used to homogeneously illuminate the detector \cite{janoschkaImplementationOperationFibercoupled2021}.
In our URHEED setup, however, the single magnetic lens was dimensioned for a small electron focal point on the detector and is unsuitable for strong de-focusing of the beam.
To solve this problem, we have mounted an additional electron disc emitter (Kimball Physics ES-535), which can be brought to the sample position from the side by means of a linear translation feedthrough (VAb LDK40-150).
We operate the emitter at 2.96\,A biased at high negative voltage of 11\,kV to maximum 20 kV in order to expose the detector with high-energy electrons.
Because the emitted currents from such cathodes can be quite substantial, the emission X-rays has to be monitored and controlled by e.g. lead glass covers on the chamber windows.
At cathode voltages exceeding 20 kV, X-ray fluorescence could pose an additional challenge.
The cathode used in this work has an activation layer to reduce its work function.
The electron beam emerging from this setup is purely divergent and thus delivers a diffuse, magnified image of the cathode surface.
Because of the granular structure of the activation layer, the beam has therefore an inhomogeneous emission profile.
To furthermore improve the homogeneity of the flat image we averaged out any spatial inhomogeneity of the emitted electrons by deflection through AC magnetic fields. Two pairs of parallel coils were placed at the outside of the chamber between the disk emitter and the gate valve for deflection in horizontal and vertical  direction.
The coils were fed with AC current generated from a function generator (Joy-IT JDS6600) and amplified with a reference audio amplifier (Behringer A800).
All relevant parameters are summarized in TABLE \ref{tab:coil_Parameter}.
While at constant electron emission from the disc source, we varied the exposure time in the range from 100\,ms to 1000\,ms in order to take flat images with different mean intensities.
The resulting flat images only exhibit a maximum intensity inhomogeneity of $\pm10$\,\%, which we considered sufficient for the correction of imaging artifacts.

\begin{table}
    \centering
    \caption{\textbf{Parameters} for the magnetic coils used for smoothing and distributing the electron pattern emitted from the disc emitter to obtain homogeneous illumination for flatfielding. }
    \begin{tabularx}{1.000\columnwidth}{llcc}
                &        & horizontal & vertical\\
         \hline
        coil    & number of windings & 48 & 72\\
         \hline
        signal  & waveform & CMOS & CMOS\\
                & amplitude / V & 2.4 & 2.1\\
                & offset / V & 0.0 & 0.0 \\
                & frequency / Hz & 130 & 90 \\
                & amplification / dB & -5 & -3\\
        \hline
        measured & coil radius / cm & 15 & 13.5 \\
                 & interaction length / cm & 15 & 11 \\
                 & maximum current / A & 2.9 & 3.6\\
        \hline
        estimated & magnetic flux / mT & $\sim$0.8 & $\sim$1.7 \\
                & deflection on detector / cm & $\pm$7 & $\pm$13\\
    \end{tabularx}
    \label{tab:coil_Parameter}
\end{table}

\begin{figure*}[htbp] 
\centering
\includegraphics[width=1.000\textwidth]{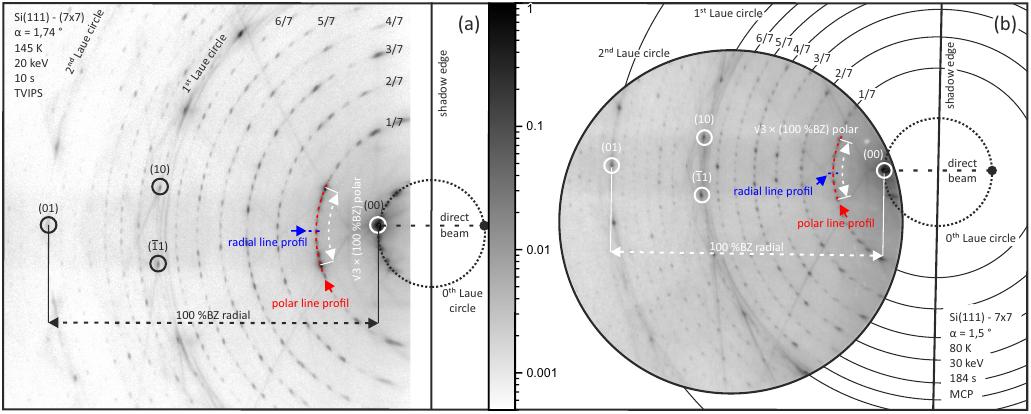} 
\caption{
\textbf{Resolution} of the TemCam compared to the MCP detector. RHEED image of the Si(111)-(7$\times$7) reconstruction taken with (a) the TemCam and (b) with the MCP-detector. Both patterns are normalized using the same logarithmic gray scale representation. The images are scaled so that the ratio of the diameters of the 0th Laue circles corresponds to the ratio of the doubled angles of incidence.
}
\label{fig:Image_compare}
\end{figure*}

\section{Detector Performance}

In order to evaluate the performance of the TemCam used in our URHEED experiment, we compare it with the previously used MCP detector \cite{hafke_pulsed_2019, janzenUltrafastElectronDiffraction2006, hanisch-blicharskiUltrafastElectronDiffraction2013, horn-vonhoegenStructuralDynamicsSurfaces2024}.
The TemCam features a square-shaped detector area with dimensions of $63.5\times63.5\,\rm{mm^2}$, comprising $4096\times4096$\,pixels and exhibiting a 16\,bit dynamic range. The MCP was imaged by a cooled CCD camera (pco.1600, PCO) with 14\,bit dynamic range and $1600\times1200$\,pixels, with the diameter of the MCP image corresponding to 750\,pixels.
Especially for diffraction experiments in which the intensities of the signals in the diffraction patterns differ by several orders of magnitude, a factor 4$\times$ larger dynamic range also allows weak signals to be measured without overexposing the strong spots.

A trivial but very valuable advantage of using such a fiber-coupled CMOS detector in a diffraction experiment is its insensitivity to overexposure by electrons. In contrast, a MCP experiences severe damage upon overexposure which easily occurs for strong diffraction spots or illumination through the direct electron beam. 
This robustness to damage renders it possible to characterize the direct beam without reducing its intensity, which otherwise could change properties like the beam focus.
With a deflector in the electron gun the electron beam can be directed onto the TemCam and only the integration time needs to be reduced such that the image of the direct beam is not overexposed. 

To further evaluate the performance of the new detector, we employed the diffraction pattern of the Si(111)-(7$\times$7) reconstruction as a reference system. An image of the pattern taken with the TemCam is shown in FIG.~\ref{fig:Image_compare} (a)).
The exposure time was 10\,s, the dark and flatcorrection was applied and it was taken with a 500\,$\rm{\mu m}$ sized electron spot  at 20\,keV under an incidence angle of 1.74\textdegree and at a sample temperature of 145\,K.

For comparison, FIG.~\ref{fig:Image_compare} (b)) displays a diffraction pattern taken from the Si(111)-(7$\times$7) reconstruction using the MCP detector unit\cite{hafke_pulsed_2019}.
This image was acquired through summing up 92 exposures, each 2\,s long, and taken with a 310\,$\rm{\mu m}$ sized electron spot at 30\,keV under an incidence angle of 1.5\textdegree at a sample temperature of 80\,K. The minimum quality diameter of the circular MCP is 40\,mm with a channel diameter of 10\,$\rm{\mu m}$ and center-to-center spacing of 12\,$\rm{\mu m}$ .

Both diffraction patterns can be compared directly as they are normalized to their maximum intensit, so they share the same logarithmic gray scale representation and display the same scale in (reciprocal) angle space. Due to the 1.79$\times$ larger detection area, the TemCam covers a much larger fraction of the diffraction pattern and more spots become visible. 

\begin{figure}[htbp]
\centering
\includegraphics[width=1.000\columnwidth]{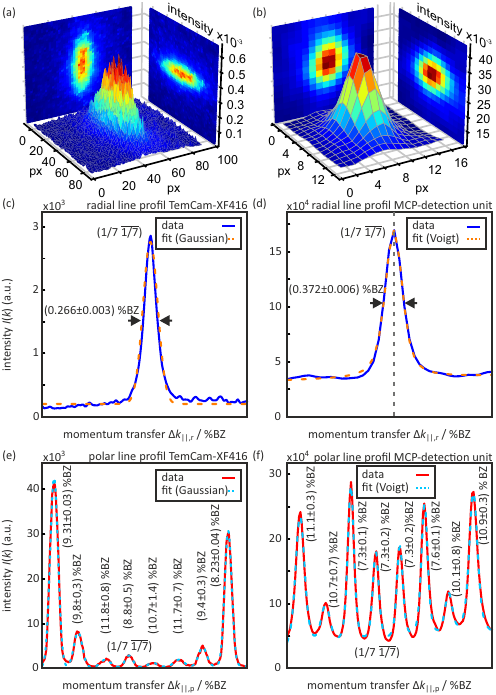} 
\caption{
\textbf{Resolution.} The (1/7 $\overline{1/7}$) spot is plotted in 3-D below the corresponding images  \ref{fig:Image_compare}in (a) and (b). The blue line-profiles (c) \& (d) are taken in radial direction from this (1/7 $\overline{1/7}$) spots, the red line-profiles (e) \& (f) in polar direction along the 1/7\,th Laue circle as marked in blue and red in both RHEED images.
}
\label{fig:Resolution}
\end{figure}

Due to the higher pixel density each diffraction spot was imaged with a larger number of pixels. 
For the MCP and for the TemCam each pixel of the image corresponds to 2844\,$\rm{\mu m^2}$ and 240\,$\rm{\mu m^2}$, respectively.
This difference of a factor of twelve in the detector resolution is clearly visible in FIG.~\ref{fig:Resolution}.
The (1/7 $\overline{1/7}$) spot of each of the two patterns is shown in (a) and (b): the difference is remarkable. 
The elongated elliptical shape of the spot, as measured with the new detector, was not resolvable with the MCP due to the limited number of pixels and the inherent blooming effect \cite{richterPositionsensitiveDetectorPerformance1986}. 
Intensity line profiles in radial FIG.~\ref{fig:Resolution}(c) \& (d) and polar (e) \& (f) direction enable a comparison of the resolution in reciprocal space in both directions. The scale of the axes were calibrated using the distance of first-order spots in each direction and corrected for the non-linearity of the RHEED geometry.

For the new detector a Gaussian profile with a FWHM of $(0.266 \pm 0.003)$\,\%BZ describes the profile in radial direction better. 
For the MCP detector a Voigt profile with a FWHM of $(0.372 \pm 0.006)$\,\%BZ is more suitable and is explained by the lower number of pixels and the blooming effect.
The FWHM corresponds to a transverse coherence lengths in radial of $\xi_{\rm{r,TemCam}} = (125 \pm 1)$\,nm for the TemCam and $\xi_{\rm{r,MCP}} = (89 \pm 1)$\,nm, for the MCP detector.

The intensity profiles through a series of (7$\times$7) spots along the polar direction are shown in FIG.~\ref{fig:Resolution}((e) \& (f)). The difference in relative intensities is explained by the different scattering conditions, i.e., different angle of incidence and energy of the electrons ($\Delta\alpha = 0.27$\textdegree).
Nevertheless, the FWHM of all spots are in a comparable regime.
For the (1/7 $\overline{1/7}$) spot the resulting transverse coherence lengths in polar direction are $\xi_{\rm{p,TemCam}} = (3.8 \pm 0.2)$\,nm for the TemCam and $\xi_{\rm{p,MCP}} = (4.5 \pm 0.1)$\,nm, for the MCP detector.
The detrimental influence of the low number of pixels and the blooming effect of the MCP detector becomes obvious along the polar direction: the wings of all spots merge into one another. 

Even after fitting Voigt profiles to the spots a residual background of 0.4\,\% of the maximum spot intensity remains. This high background level originates from camera-related thermally induced background which has been subtracted -- the level of shot noise, however, still contributes to the diffraction image. As a second contribution the inherent high noise level of the MCP detector is not subtracted, because no dark images are available.
For the TemCam the residual background is only 0.1\,\% of the maximum spot intensity.

With the TemCam all (7$\times$7) spots can be described well with Gaussian profiles. As no blooming effect is present, the wings of the spots decay without merging into one another.
While a quantitative comparison of the relative intensities is difficult due to the different scattering conditions, we observe a qualitative difference in signal-to-background ratio of 530:1 and 12:1 for the TemCam and MCP detector, respectively.

\begin{figure}[htbp]
\centering
\includegraphics[width=1.000\columnwidth]{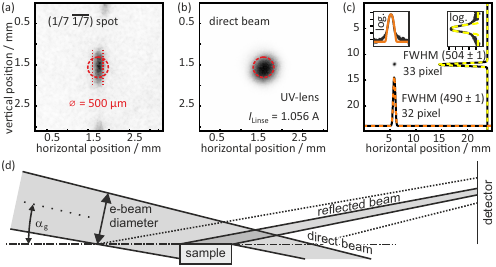} 
\caption{
\textbf{Spot shape}
(a) The (1/7 $\overline{1/7}$) spot from the image taken with the TemCam displayed in units of mm as imaged on the camera.
(b) Spot of the focused direct beam, imaged with 100\,ms exposure time. Same scale as in (a). 
(c) Direct beam, zoomed out by a factor of eight. Orange and yellow dotted lines are Gaussian curves fitted to the horizontal and vertical line profiles, respectively. Small insets display them in logarithmic scale to show the good agreement. (d) Scheme of the beam path illustrating that the sample width limits the angular distribution of the electron beam in one dimension resulting in a higher radial resolution.
}
\label{fig:Sample_as_slit}
\end{figure}

Inspection of the spots on the 1/7 Laue ring at higher magnification (see FIG.~\ref{fig:Sample_as_slit}(a)) reveals a surprising finding.
The (7$\times$7) spots exhibit a larger width along the polar direction as compared to the width along the radial direction. 
For diffraction from a perfect surface one would expect a circular spot \cite{henzlerMeasurementSurfaceDefects1984a} at a width given by the focus properties of the electron gun -- for the (00)-spot in RHEED the sample principally acts as a mirror. 
However, due to defects at the surface like steps, domain boundaries, etc. all spots experience additional broadening \cite{horn-vonhoegenGrowthSemiconductorLayers1999, kleinLostReciprocalSpace2011}. 
Given the higher transverse coherence length along the radial direction, the broadening should be more pronounced in this direction. 
Here, however, we observe the opposite!

The direct electron beam is shown in FIG.~\ref{fig:Sample_as_slit}((b) \& (c)) and exhibits the expected Gaussian-shaped circular spot profile with a width which agrees with the spot width in polar direction. 
In general, the width of the direct beam is determined by the electron source size at the photocathode, the magnification during imaging onto the detector, and spherical aberrations of the single magnetic lens \cite{hafke_pulsed_2019}.

We attribute the smaller spot width in the radial direction to the cancellation/suppression of off-axis beams that experience large spherical aberrations from the image construction. 

In RHEED, the grazing incidence of 1-6\textdegree  ~ensures surface sensitivity \cite{AppliedRHEED1999}. 
Here, the electron beam is incident at a grazing angle of $ \alpha_{\rm{g}} = 1.74$\textdegree ~with respect to the sample surface. 
This results in a so-called foreshortening factor of $1/\sin(\alpha_{\rm{g}})\simeq 33$, i.e., the beam size on the sample position is elongated by a factor of 33.

Thus, for a finite beam diameter of $\simeq 300 ~\mu$m the sample acts as a narrow slit of 1.7~mm/33 = 0.05~mm width for the image construction of the diffraction pattern as sketched in FIG.~\ref{fig:Sample_as_slit}(d). 
In radial direction only the central on-axis beams contribute to image formation. 
The off-axis beams that experience large spherical aberrations are suppressed from image formation, because they are either blocked or do not hit the sample. 
Consequently, the spots become sharper in this direction with a superior transverse coherence length of $\xi_{\rm{r,TemCam}} = (125 \pm 1)$\,nm. 

\section{Conclusions}
We have demonstrated that with the help of a water-cooled cooling collar and a differential pumpable gate valve the bake-out of temperature-sensitive equipment like the TemCam is possible without dismounting the camera. 

By this upgrade from a MCP detector to the TemCam  camera we improved the resolution in reciprocal space by 30\,\%. Thus, we are able to perform spot profile analysis without MCP-related blooming effects and at nine times higher pixel density.
The low background intensity of the TemCam camera enables to study weak features in the diffraction pattern such as Kikuchi lines and bands that were previously inaccessible in our URHEED experiment. 

One drawback we have to accept for measurements with the TemCam at full resolution is the inherent dead time of 1.2\,s due to the readout of the 4k CMOS chip.
For adjustment and optimization, however, the detector is fast enough at 10\,Hz in continuous rolling shutter mode at full resolution.
Also the 4$\times$ higher dynamic range allow for measuring analyzable data of weak and strong diffraction signals simultaneously which reduces the overall measuring time.

Due to its scintillator/CMOS chip-based design, the TemCam is inherently sensitive to photons with an energy above the indirect band gap of Si.
This must be taken into account when the TemCam is employed in experiments with optical excitation. When pumping with wavelength of 800\,nm photons from a Ti:sapphire laser, effective shielding of the exciting pump laser pulse must be considered. Alternatively, photons with an energy $h\nu < E_{\rm{gap}}=1.12$\,eV can be used as pump pulse.
The light sensitivity is reduced for setups using higher electron energies. Then the Al layer covering the scintillator is thicker than the 30 nm used for the TemCam used in our case which is optimized for the 30\,keV electrons employed in our URHEED experiment.

A major advantage in daily operation of the TemCam, however, is its insensitivity and robustness to destruction caused by overexposure in the case of intense diffraction spots or by the direct electron beam -- thus avoiding imprinting the groups unique damage pattern into each diffraction image!

\section*{Acknowledgments}
Fruitful discussions with H. R. Tietz and M. Oster from TVIPS and with R. Ernstdorfer, P. Baum, G. Sciaini, P. Dreher, A. Neuhaus, and F.-J. Meyer zu Heringdorf are gratefully acknowledged. Funded by the Deutsche Forschungsgemeinschaft (DFG, German Research Foundation) through project C03 of Collaborative Research Center SFB1242 ``Nonequilibrium dynamics of condensed matter in the time domain'' (Project-ID 278162697).

\section*{Author Declaration}

\subsection{Conflict of Interest}
The authors have no conflicts to disclose.

\subsection{Author Contributions}
\textbf{Jonas D. Fortmann}:
Conceptualization (lead);
Data curation (equal);
Visualization (lead);
Formal analysis (equal);
Writing – original draft (lead);
Writing – review \& editing (equal).
\textbf{Alexander Kaßen}:
Data curation (equal)
Formal analysis (equal)
Visualization (supporting);
\textbf{Christian Brand}:
Conceptualization (supporting);
Construction (equal);
Writing – review \& editing (supporting).
\textbf{Thomas Duden}:
Construction (equal);
Writing – review \& editing (supporting).
\textbf{Michael Horn-von Hoegen}:
Conceptualization (supporting);
Supervision (lead); 
Writing – original draft (supporting);
Writing – review \& editing (equal).

\section*{Data Availability}

The data that supports the findings of this study are available from the corresponding author upon reasonable request.

\section*{Appendix A: Implementation Image}

\section*{References}

\bibliography{main}

\begin{thebibliography}{31}%
\makeatletter
\providecommand \@ifxundefined [1]{%
 \@ifx{#1\undefined}
}%
\providecommand \@ifnum [1]{%
 \ifnum #1\expandafter \@firstoftwo
 \else \expandafter \@secondoftwo
 \fi
}%
\providecommand \@ifx [1]{%
 \ifx #1\expandafter \@firstoftwo
 \else \expandafter \@secondoftwo
 \fi
}%
\providecommand \natexlab [1]{#1}%
\providecommand \enquote  [1]{``#1''}%
\providecommand \bibnamefont  [1]{#1}%
\providecommand \bibfnamefont [1]{#1}%
\providecommand \citenamefont [1]{#1}%
\providecommand \href@noop [0]{\@secondoftwo}%
\providecommand \href [0]{\begingroup \@sanitize@url \@href}%
\providecommand \@href[1]{\@@startlink{#1}\@@href}%
\providecommand \@@href[1]{\endgroup#1\@@endlink}%
\providecommand \@sanitize@url [0]{\catcode `\\12\catcode `\$12\catcode `\&12\catcode `\#12\catcode `\^12\catcode `\_12\catcode `\%12\relax}%
\providecommand \@@startlink[1]{}%
\providecommand \@@endlink[0]{}%
\providecommand \url  [0]{\begingroup\@sanitize@url \@url }%
\providecommand \@url [1]{\endgroup\@href {#1}{\urlprefix }}%
\providecommand \urlprefix  [0]{URL }%
\providecommand \Eprint [0]{\href }%
\providecommand \doibase [0]{http://dx.doi.org/}%
\providecommand \selectlanguage [0]{\@gobble}%
\providecommand \bibinfo  [0]{\@secondoftwo}%
\providecommand \bibfield  [0]{\@secondoftwo}%
\providecommand \translation [1]{[#1]}%
\providecommand \BibitemOpen [0]{}%
\providecommand \bibitemStop [0]{}%
\providecommand \bibitemNoStop [0]{.\EOS\space}%
\providecommand \EOS [0]{\spacefactor3000\relax}%
\providecommand \BibitemShut  [1]{\csname bibitem#1\endcsname}%
\let\auto@bib@innerbib\@empty
\bibitem [{\citenamefont {Niu}\ \emph {et~al.}(2023)\citenamefont {Niu}, \citenamefont {Vinogradov}, \citenamefont {Preobrajenski}, \citenamefont {Struzzi}, \citenamefont {Sarpi}, \citenamefont {Zhu}, \citenamefont {Golias},\ and\ \citenamefont {Zakharov}}]{niuMAXPEEMSpectromicroscopyBeamline2023}%
  \BibitemOpen
  \bibfield  {author} {\bibinfo {author} {\bibfnamefont {Y.}~\bibnamefont {Niu}}, \bibinfo {author} {\bibfnamefont {N.}~\bibnamefont {Vinogradov}}, \bibinfo {author} {\bibfnamefont {A.}~\bibnamefont {Preobrajenski}}, \bibinfo {author} {\bibfnamefont {C.}~\bibnamefont {Struzzi}}, \bibinfo {author} {\bibfnamefont {B.}~\bibnamefont {Sarpi}}, \bibinfo {author} {\bibfnamefont {L.}~\bibnamefont {Zhu}}, \bibinfo {author} {\bibfnamefont {E.}~\bibnamefont {Golias}}, \ and\ \bibinfo {author} {\bibfnamefont {A.}~\bibnamefont {Zakharov}},\ }\bibfield  {title} {\enquote {\bibinfo {title} {{{MAXPEEM}}: A spectromicroscopy beamline at {{MAX IV}} laboratory},}\ }\href {\doibase 10.1107/S160057752300019X} {\bibfield  {journal} {\bibinfo  {journal} {Journal of Synchrotron Radiation}\ }\textbf {\bibinfo {volume} {30}},\ \bibinfo {pages} {468--478} (\bibinfo {year} {2023})}\BibitemShut {NoStop}%
\bibitem [{\citenamefont {Janoschka}\ \emph {et~al.}(2021)\citenamefont {Janoschka}, \citenamefont {Dreher}, \citenamefont {R{\"o}dl}, \citenamefont {Franz}, \citenamefont {Schaff}, \citenamefont {{Horn-von Hoegen}},\ and\ \citenamefont {{Meyer zu Heringdorf}}}]{janoschkaImplementationOperationFibercoupled2021}%
  \BibitemOpen
  \bibfield  {author} {\bibinfo {author} {\bibfnamefont {D.}~\bibnamefont {Janoschka}}, \bibinfo {author} {\bibfnamefont {P.}~\bibnamefont {Dreher}}, \bibinfo {author} {\bibfnamefont {A.}~\bibnamefont {R{\"o}dl}}, \bibinfo {author} {\bibfnamefont {T.}~\bibnamefont {Franz}}, \bibinfo {author} {\bibfnamefont {O.}~\bibnamefont {Schaff}}, \bibinfo {author} {\bibfnamefont {M.}~\bibnamefont {{Horn-von Hoegen}}}, \ and\ \bibinfo {author} {\bibfnamefont {F.~J.}\ \bibnamefont {{Meyer zu Heringdorf}}},\ }\bibfield  {title} {\enquote {\bibinfo {title} {Implementation and operation of a fiber-coupled {{CMOS}} detector in a low energy electron {{Microscope}}},}\ }\href {\doibase 10.1016/j.ultramic.2020.113180} {\bibfield  {journal} {\bibinfo  {journal} {Ultramicroscopy}\ }\textbf {\bibinfo {volume} {221}},\ \bibinfo {pages} {113180} (\bibinfo {year} {2021})}\BibitemShut {NoStop}%
\bibitem [{\citenamefont {Waldecker}\ \emph {et~al.}(2015)\citenamefont {Waldecker}, \citenamefont {Miller}, \citenamefont {Rud{\'e}}, \citenamefont {Bertoni}, \citenamefont {Osmond}, \citenamefont {Pruneri}, \citenamefont {Simpson}, \citenamefont {Ernstorfer},\ and\ \citenamefont {Wall}}]{waldeckerTimedomainSeparationOptical2015}%
  \BibitemOpen
  \bibfield  {author} {\bibinfo {author} {\bibfnamefont {L.}~\bibnamefont {Waldecker}}, \bibinfo {author} {\bibfnamefont {T.~A.}\ \bibnamefont {Miller}}, \bibinfo {author} {\bibfnamefont {M.}~\bibnamefont {Rud{\'e}}}, \bibinfo {author} {\bibfnamefont {R.}~\bibnamefont {Bertoni}}, \bibinfo {author} {\bibfnamefont {J.}~\bibnamefont {Osmond}}, \bibinfo {author} {\bibfnamefont {V.}~\bibnamefont {Pruneri}}, \bibinfo {author} {\bibfnamefont {R.~E.}\ \bibnamefont {Simpson}}, \bibinfo {author} {\bibfnamefont {R.}~\bibnamefont {Ernstorfer}}, \ and\ \bibinfo {author} {\bibfnamefont {S.}~\bibnamefont {Wall}},\ }\bibfield  {title} {\enquote {\bibinfo {title} {Time-domain separation of optical properties from structural transitions in resonantly bonded materials},}\ }\href {\doibase 10.1038/nmat4359} {\bibfield  {journal} {\bibinfo  {journal} {Nature Materials}\ }\textbf {\bibinfo {volume} {14}},\ \bibinfo {pages} {991--995} (\bibinfo {year} {2015})}\BibitemShut {NoStop}%
\bibitem [{\citenamefont {Grob}\ \emph {et~al.}(2013)\citenamefont {Grob}, \citenamefont {Bean}, \citenamefont {Typke}, \citenamefont {Li}, \citenamefont {Nogales},\ and\ \citenamefont {Glaeser}}]{grobRankingTEMCameras2013}%
  \BibitemOpen
  \bibfield  {author} {\bibinfo {author} {\bibfnamefont {P.}~\bibnamefont {Grob}}, \bibinfo {author} {\bibfnamefont {D.}~\bibnamefont {Bean}}, \bibinfo {author} {\bibfnamefont {D.}~\bibnamefont {Typke}}, \bibinfo {author} {\bibfnamefont {X.}~\bibnamefont {Li}}, \bibinfo {author} {\bibfnamefont {E.}~\bibnamefont {Nogales}}, \ and\ \bibinfo {author} {\bibfnamefont {R.~M.}\ \bibnamefont {Glaeser}},\ }\bibfield  {title} {\enquote {\bibinfo {title} {Ranking {{TEM}} cameras by their response to electron shot noise},}\ }\href {\doibase 10.1016/j.ultramic.2013.01.003} {\bibfield  {journal} {\bibinfo  {journal} {Ultramicroscopy}\ }\textbf {\bibinfo {volume} {133}},\ \bibinfo {pages} {1--7} (\bibinfo {year} {2013})}\BibitemShut {NoStop}%
\bibitem [{\citenamefont {Tyson}\ \emph {et~al.}(2024)\citenamefont {Tyson}, \citenamefont {Amarasinghe}, \citenamefont {Abeykoon}, \citenamefont {Lalancette}, \citenamefont {Du}, \citenamefont {Fang}, \citenamefont {Cheong}, \citenamefont {{Al-Mahboob}},\ and\ \citenamefont {Sadowski}}]{tysonSurfaceMagnetismFe3GeTe22024}%
  \BibitemOpen
  \bibfield  {author} {\bibinfo {author} {\bibfnamefont {T.~A.}\ \bibnamefont {Tyson}}, \bibinfo {author} {\bibfnamefont {S.}~\bibnamefont {Amarasinghe}}, \bibinfo {author} {\bibfnamefont {A.~M.}\ \bibnamefont {Abeykoon}}, \bibinfo {author} {\bibfnamefont {R.}~\bibnamefont {Lalancette}}, \bibinfo {author} {\bibfnamefont {S.~K.}\ \bibnamefont {Du}}, \bibinfo {author} {\bibfnamefont {X.}~\bibnamefont {Fang}}, \bibinfo {author} {\bibfnamefont {S.-W.}\ \bibnamefont {Cheong}}, \bibinfo {author} {\bibfnamefont {A.}~\bibnamefont {{Al-Mahboob}}}, \ and\ \bibinfo {author} {\bibfnamefont {J.~T.}\ \bibnamefont {Sadowski}},\ }\href {\doibase 10.48550/arXiv.2409.03565} {\enquote {\bibinfo {title} {Surface {{Magnetism}} in {{Fe3GeTe2 Crystals}}},}\ } (\bibinfo {year} {2024}),\ \Eprint {http://arxiv.org/abs/2409.03565} {arXiv:2409.03565} \BibitemShut {NoStop}%
\bibitem [{\citenamefont {Ruskin}, \citenamefont {Yu},\ and\ \citenamefont {Grigorieff}(2013)}]{ruskinQuantitativeCharacterizationElectron2013}%
  \BibitemOpen
  \bibfield  {author} {\bibinfo {author} {\bibfnamefont {A.~I.}\ \bibnamefont {Ruskin}}, \bibinfo {author} {\bibfnamefont {Z.}~\bibnamefont {Yu}}, \ and\ \bibinfo {author} {\bibfnamefont {N.}~\bibnamefont {Grigorieff}},\ }\bibfield  {title} {\enquote {\bibinfo {title} {Quantitative characterization of electron detectors for transmission electron microscopy},}\ }\href {\doibase 10.1016/j.jsb.2013.10.016} {\bibfield  {journal} {\bibinfo  {journal} {Journal of Structural Biology}\ }\textbf {\bibinfo {volume} {184}},\ \bibinfo {pages} {385--393} (\bibinfo {year} {2013})}\BibitemShut {NoStop}%
\bibitem [{\citenamefont {Rauch}\ \emph {et~al.}(2021)\citenamefont {Rauch}, \citenamefont {Harrison}, \citenamefont {Zhou}, \citenamefont {Herbig}, \citenamefont {Ludwig},\ and\ \citenamefont {V{\'e}ron}}]{rauchNewFeaturesCrystal2021}%
  \BibitemOpen
  \bibfield  {author} {\bibinfo {author} {\bibfnamefont {E.~F.}\ \bibnamefont {Rauch}}, \bibinfo {author} {\bibfnamefont {P.}~\bibnamefont {Harrison}}, \bibinfo {author} {\bibfnamefont {X.}~\bibnamefont {Zhou}}, \bibinfo {author} {\bibfnamefont {M.}~\bibnamefont {Herbig}}, \bibinfo {author} {\bibfnamefont {W.}~\bibnamefont {Ludwig}}, \ and\ \bibinfo {author} {\bibfnamefont {M.}~\bibnamefont {V{\'e}ron}},\ }\bibfield  {title} {\enquote {\bibinfo {title} {New {{Features}} in {{Crystal Orientation}} and {{Phase Mapping}} for {{Transmission Electron Microscopy}}},}\ }\href {\doibase 10.3390/sym13091675} {\bibfield  {journal} {\bibinfo  {journal} {Symmetry}\ }\textbf {\bibinfo {volume} {13}},\ \bibinfo {pages} {1675} (\bibinfo {year} {2021})}\BibitemShut {NoStop}%
\bibitem [{\citenamefont {Henke}\ \emph {et~al.}(2021)\citenamefont {Henke}, \citenamefont {Raja}, \citenamefont {Feist}, \citenamefont {Huang}, \citenamefont {Arend}, \citenamefont {Yang}, \citenamefont {Kappert}, \citenamefont {Wang}, \citenamefont {M{\"o}ller}, \citenamefont {Pan}, \citenamefont {Liu}, \citenamefont {Kfir}, \citenamefont {Ropers},\ and\ \citenamefont {Kippenberg}}]{henkeIntegratedPhotonicsEnables2021}%
  \BibitemOpen
  \bibfield  {author} {\bibinfo {author} {\bibfnamefont {J.-W.}\ \bibnamefont {Henke}}, \bibinfo {author} {\bibfnamefont {A.~S.}\ \bibnamefont {Raja}}, \bibinfo {author} {\bibfnamefont {A.}~\bibnamefont {Feist}}, \bibinfo {author} {\bibfnamefont {G.}~\bibnamefont {Huang}}, \bibinfo {author} {\bibfnamefont {G.}~\bibnamefont {Arend}}, \bibinfo {author} {\bibfnamefont {Y.}~\bibnamefont {Yang}}, \bibinfo {author} {\bibfnamefont {F.~J.}\ \bibnamefont {Kappert}}, \bibinfo {author} {\bibfnamefont {R.~N.}\ \bibnamefont {Wang}}, \bibinfo {author} {\bibfnamefont {M.}~\bibnamefont {M{\"o}ller}}, \bibinfo {author} {\bibfnamefont {J.}~\bibnamefont {Pan}}, \bibinfo {author} {\bibfnamefont {J.}~\bibnamefont {Liu}}, \bibinfo {author} {\bibfnamefont {O.}~\bibnamefont {Kfir}}, \bibinfo {author} {\bibfnamefont {C.}~\bibnamefont {Ropers}}, \ and\ \bibinfo {author} {\bibfnamefont {T.~J.}\ \bibnamefont {Kippenberg}},\ }\bibfield  {title} {\enquote {\bibinfo {title} {Integrated photonics enables continuous-beam electron phase
  modulation},}\ }\href {\doibase 10.1038/s41586-021-04197-5} {\bibfield  {journal} {\bibinfo  {journal} {Nature}\ }\textbf {\bibinfo {volume} {600}},\ \bibinfo {pages} {653--658} (\bibinfo {year} {2021})}\BibitemShut {NoStop}%
\bibitem [{\citenamefont {Tak}\ \emph {et~al.}(2023)\citenamefont {Tak}, \citenamefont {Johnson}, \citenamefont {Ho}, \citenamefont {Wu}, \citenamefont {Sauty}, \citenamefont {Rebollo}, \citenamefont {Schmid}, \citenamefont {Peretti}, \citenamefont {Wu}, \citenamefont {Weisbuch},\ and\ \citenamefont {Speck}}]{takInjectionMechanismsIIInitride2023}%
  \BibitemOpen
  \bibfield  {author} {\bibinfo {author} {\bibfnamefont {T.}~\bibnamefont {Tak}}, \bibinfo {author} {\bibfnamefont {C.~W.}\ \bibnamefont {Johnson}}, \bibinfo {author} {\bibfnamefont {W.~Y.}\ \bibnamefont {Ho}}, \bibinfo {author} {\bibfnamefont {F.}~\bibnamefont {Wu}}, \bibinfo {author} {\bibfnamefont {M.}~\bibnamefont {Sauty}}, \bibinfo {author} {\bibfnamefont {S.}~\bibnamefont {Rebollo}}, \bibinfo {author} {\bibfnamefont {A.~K.}\ \bibnamefont {Schmid}}, \bibinfo {author} {\bibfnamefont {J.}~\bibnamefont {Peretti}}, \bibinfo {author} {\bibfnamefont {Y.-R.}\ \bibnamefont {Wu}}, \bibinfo {author} {\bibfnamefont {C.}~\bibnamefont {Weisbuch}}, \ and\ \bibinfo {author} {\bibfnamefont {J.~S.}\ \bibnamefont {Speck}},\ }\bibfield  {title} {\enquote {\bibinfo {title} {Injection mechanisms in a {{III-nitride}} light-emitting diode as seen by self-emissive electron microscopy},}\ }\href {\doibase 10.1103/PhysRevApplied.20.064045} {\bibfield  {journal} {\bibinfo  {journal} {Physical Review Applied}\ }\textbf {\bibinfo
  {volume} {20}},\ \bibinfo {pages} {064045} (\bibinfo {year} {2023})}\BibitemShut {NoStop}%
\bibitem [{\citenamefont {Mattes}, \citenamefont {Volkov},\ and\ \citenamefont {Baum}(2024)}]{mattesFemtosecondElectronBeam2024}%
  \BibitemOpen
  \bibfield  {author} {\bibinfo {author} {\bibfnamefont {M.}~\bibnamefont {Mattes}}, \bibinfo {author} {\bibfnamefont {M.}~\bibnamefont {Volkov}}, \ and\ \bibinfo {author} {\bibfnamefont {P.}~\bibnamefont {Baum}},\ }\bibfield  {title} {\enquote {\bibinfo {title} {Femtosecond electron beam probe of ultrafast electronics},}\ }\href {\doibase 10.1038/s41467-024-45744-8} {\bibfield  {journal} {\bibinfo  {journal} {Nature Communications}\ }\textbf {\bibinfo {volume} {15}},\ \bibinfo {pages} {1743} (\bibinfo {year} {2024})}\BibitemShut {NoStop}%
\bibitem [{\citenamefont {Faruqi}\ and\ \citenamefont {McMullan}(2011)}]{faruqiElectronicDetectorsElectron2011}%
  \BibitemOpen
  \bibfield  {author} {\bibinfo {author} {\bibfnamefont {A.~R.}\ \bibnamefont {Faruqi}}\ and\ \bibinfo {author} {\bibfnamefont {G.}~\bibnamefont {McMullan}},\ }\bibfield  {title} {\enquote {\bibinfo {title} {Electronic detectors for electron microscopy},}\ }\href {\doibase 10.1017/S0033583511000035} {\bibfield  {journal} {\bibinfo  {journal} {Quarterly Reviews of Biophysics}\ }\textbf {\bibinfo {volume} {44}},\ \bibinfo {pages} {357--390} (\bibinfo {year} {2011})}\BibitemShut {NoStop}%
\bibitem [{\citenamefont {Weber}\ and\ \citenamefont {Sch{\"a}fer}(2024)}]{weberElectronImagingNanoscale2024}%
  \BibitemOpen
  \bibfield  {author} {\bibinfo {author} {\bibfnamefont {J.~T.}\ \bibnamefont {Weber}}\ and\ \bibinfo {author} {\bibfnamefont {S.}~\bibnamefont {Sch{\"a}fer}},\ }\bibfield  {title} {\enquote {\bibinfo {title} {Electron {{Imaging}} of {{Nanoscale Charge Distributions Induced}} by {{Femtosecond Light Pulses}}},}\ }\href {\doibase 10.1021/acs.nanolett.4c00773} {\bibfield  {journal} {\bibinfo  {journal} {Nano Letters}\ }\textbf {\bibinfo {volume} {24}},\ \bibinfo {pages} {5746--5753} (\bibinfo {year} {2024})}\BibitemShut {NoStop}%
\bibitem [{\citenamefont {Hugenschmidt}\ \emph {et~al.}(2021)\citenamefont {Hugenschmidt}, \citenamefont {Kutonova}, \citenamefont {S{\'a}nchez}, \citenamefont {Moulai}, \citenamefont {Gliemann}, \citenamefont {Br{\"a}se}, \citenamefont {W{\"o}ll},\ and\ \citenamefont {Gerthsen}}]{hugenschmidtDirectSynthesisZIF82021}%
  \BibitemOpen
  \bibfield  {author} {\bibinfo {author} {\bibfnamefont {M.}~\bibnamefont {Hugenschmidt}}, \bibinfo {author} {\bibfnamefont {K.}~\bibnamefont {Kutonova}}, \bibinfo {author} {\bibfnamefont {E.~P.~V.}\ \bibnamefont {S{\'a}nchez}}, \bibinfo {author} {\bibfnamefont {S.}~\bibnamefont {Moulai}}, \bibinfo {author} {\bibfnamefont {H.}~\bibnamefont {Gliemann}}, \bibinfo {author} {\bibfnamefont {S.}~\bibnamefont {Br{\"a}se}}, \bibinfo {author} {\bibfnamefont {C.}~\bibnamefont {W{\"o}ll}}, \ and\ \bibinfo {author} {\bibfnamefont {D.}~\bibnamefont {Gerthsen}},\ }\bibfield  {title} {\enquote {\bibinfo {title} {Direct {{Synthesis}} of {{ZIF-8}} on {{Transmission Electron Microscopy Grids Allows Structure Analysis}} and {{3D Reconstruction}}},}\ }\href {\doibase 10.1017/S1431927621010771} {\bibfield  {journal} {\bibinfo  {journal} {Microscopy and Microanalysis}\ }\textbf {\bibinfo {volume} {27}},\ \bibinfo {pages} {3114--3116} (\bibinfo {year} {2021})}\BibitemShut {NoStop}%
\bibitem [{\citenamefont {Minenkov}\ \emph {et~al.}(2022)\citenamefont {Minenkov}, \citenamefont {{\v S}anti{\'c}}, \citenamefont {Truglas}, \citenamefont {Aberl}, \citenamefont {Vuku{\v s}i{\'c}}, \citenamefont {Brehm},\ and\ \citenamefont {Groiss}}]{minenkovAdvancedPreparationPlanview2022}%
  \BibitemOpen
  \bibfield  {author} {\bibinfo {author} {\bibfnamefont {A.}~\bibnamefont {Minenkov}}, \bibinfo {author} {\bibfnamefont {N.}~\bibnamefont {{\v S}anti{\'c}}}, \bibinfo {author} {\bibfnamefont {T.}~\bibnamefont {Truglas}}, \bibinfo {author} {\bibfnamefont {J.}~\bibnamefont {Aberl}}, \bibinfo {author} {\bibfnamefont {L.}~\bibnamefont {Vuku{\v s}i{\'c}}}, \bibinfo {author} {\bibfnamefont {M.}~\bibnamefont {Brehm}}, \ and\ \bibinfo {author} {\bibfnamefont {H.}~\bibnamefont {Groiss}},\ }\bibfield  {title} {\enquote {\bibinfo {title} {Advanced preparation of plan-view specimens on a {{MEMS}} chip for in situ {{TEM}} heating experiments},}\ }\href {\doibase 10.1557/s43577-021-00255-5} {\bibfield  {journal} {\bibinfo  {journal} {MRS Bulletin}\ }\textbf {\bibinfo {volume} {47}},\ \bibinfo {pages} {359--370} (\bibinfo {year} {2022})}\BibitemShut {NoStop}%
\bibitem [{\citenamefont {Stumpf}\ \emph {et~al.}(2010)\citenamefont {Stumpf}, \citenamefont {Bobolas}, \citenamefont {Daberkow}, \citenamefont {Fanderl}, \citenamefont {Heike}, \citenamefont {Huber}, \citenamefont {Kofler}, \citenamefont {Maniette},\ and\ \citenamefont {Tietz}}]{stumpfDesignCharacterization162010}%
  \BibitemOpen
  \bibfield  {author} {\bibinfo {author} {\bibfnamefont {M.}~\bibnamefont {Stumpf}}, \bibinfo {author} {\bibfnamefont {K.}~\bibnamefont {Bobolas}}, \bibinfo {author} {\bibfnamefont {I.}~\bibnamefont {Daberkow}}, \bibinfo {author} {\bibfnamefont {U.}~\bibnamefont {Fanderl}}, \bibinfo {author} {\bibfnamefont {T.}~\bibnamefont {Heike}}, \bibinfo {author} {\bibfnamefont {T.}~\bibnamefont {Huber}}, \bibinfo {author} {\bibfnamefont {C.}~\bibnamefont {Kofler}}, \bibinfo {author} {\bibfnamefont {Y.}~\bibnamefont {Maniette}}, \ and\ \bibinfo {author} {\bibfnamefont {H.~R.}\ \bibnamefont {Tietz}},\ }\bibfield  {title} {\enquote {\bibinfo {title} {Design and {{Characterization}} of 16 {{MegaPixel Fiber Optic Coupled CMOS Detector}} for {{Transmission Electron Microscopy}}},}\ }\href {\doibase 10.1017/S1431927610054516} {\bibfield  {journal} {\bibinfo  {journal} {Microscopy and Microanalysis}\ }\textbf {\bibinfo {volume} {16}},\ \bibinfo {pages} {856--857} (\bibinfo {year} {2010})}\BibitemShut {NoStop}%
\bibitem [{\citenamefont {Janzen}\ \emph {et~al.}(2006)\citenamefont {Janzen}, \citenamefont {Krenzer}, \citenamefont {Zhou}, \citenamefont {{von der Linde}},\ and\ \citenamefont {{Horn-von Hoegen}}}]{janzenUltrafastElectronDiffraction2006}%
  \BibitemOpen
  \bibfield  {author} {\bibinfo {author} {\bibfnamefont {A.}~\bibnamefont {Janzen}}, \bibinfo {author} {\bibfnamefont {B.}~\bibnamefont {Krenzer}}, \bibinfo {author} {\bibfnamefont {P.}~\bibnamefont {Zhou}}, \bibinfo {author} {\bibfnamefont {D.}~\bibnamefont {{von der Linde}}}, \ and\ \bibinfo {author} {\bibfnamefont {M.}~\bibnamefont {{Horn-von Hoegen}}},\ }\bibfield  {title} {\enquote {\bibinfo {title} {Ultrafast electron diffraction at surfaces after laser excitation},}\ }\href {\doibase 10.1016/j.susc.2006.02.070} {\bibfield  {journal} {\bibinfo  {journal} {Surface Science}\ }\textbf {\bibinfo {volume} {600}},\ \bibinfo {pages} {4094--4098} (\bibinfo {year} {2006})}\BibitemShut {NoStop}%
\bibitem [{\citenamefont {{Hanisch-Blicharski}}\ \emph {et~al.}(2013)\citenamefont {{Hanisch-Blicharski}}, \citenamefont {Janzen}, \citenamefont {Krenzer}, \citenamefont {Wall}, \citenamefont {Klasing}, \citenamefont {Kalus}, \citenamefont {Frigge}, \citenamefont {Kammler},\ and\ \citenamefont {{Horn-von Hoegen}}}]{hanisch-blicharskiUltrafastElectronDiffraction2013}%
  \BibitemOpen
  \bibfield  {author} {\bibinfo {author} {\bibfnamefont {A.}~\bibnamefont {{Hanisch-Blicharski}}}, \bibinfo {author} {\bibfnamefont {A.}~\bibnamefont {Janzen}}, \bibinfo {author} {\bibfnamefont {B.}~\bibnamefont {Krenzer}}, \bibinfo {author} {\bibfnamefont {S.}~\bibnamefont {Wall}}, \bibinfo {author} {\bibfnamefont {F.}~\bibnamefont {Klasing}}, \bibinfo {author} {\bibfnamefont {A.}~\bibnamefont {Kalus}}, \bibinfo {author} {\bibfnamefont {T.}~\bibnamefont {Frigge}}, \bibinfo {author} {\bibfnamefont {M.}~\bibnamefont {Kammler}}, \ and\ \bibinfo {author} {\bibfnamefont {M.}~\bibnamefont {{Horn-von Hoegen}}},\ }\bibfield  {title} {\enquote {\bibinfo {title} {Ultra-fast electron diffraction at surfaces: {{From}} nanoscale heat transport to driven phase transitions},}\ }\href {\doibase 10.1016/j.ultramic.2012.07.017} {\bibfield  {journal} {\bibinfo  {journal} {Ultramicroscopy}\ }\textbf {\bibinfo {volume} {127}},\ \bibinfo {pages} {2--8} (\bibinfo {year} {2013})}\BibitemShut {NoStop}%
\bibitem [{\citenamefont {Frigge}\ \emph {et~al.}(2017)\citenamefont {Frigge}, \citenamefont {Hafke}, \citenamefont {Witte}, \citenamefont {Krenzer}, \citenamefont {Streub{\"u}hr}, \citenamefont {Samad~Syed}, \citenamefont {Mik{\v s}i{\'c}~Trontl}, \citenamefont {Avigo}, \citenamefont {Zhou}, \citenamefont {Ligges}, \citenamefont {{von Der Linde}}, \citenamefont {Bovensiepen}, \citenamefont {{Horn-von Hoegen}}, \citenamefont {Wippermann}, \citenamefont {L{\"u}cke}, \citenamefont {Sanna}, \citenamefont {Gerstmann},\ and\ \citenamefont {Schmidt}}]{frigge_optically_2017}%
  \BibitemOpen
  \bibfield  {author} {\bibinfo {author} {\bibfnamefont {T.}~\bibnamefont {Frigge}}, \bibinfo {author} {\bibfnamefont {B.}~\bibnamefont {Hafke}}, \bibinfo {author} {\bibfnamefont {T.}~\bibnamefont {Witte}}, \bibinfo {author} {\bibfnamefont {B.}~\bibnamefont {Krenzer}}, \bibinfo {author} {\bibfnamefont {C.}~\bibnamefont {Streub{\"u}hr}}, \bibinfo {author} {\bibfnamefont {A.}~\bibnamefont {Samad~Syed}}, \bibinfo {author} {\bibfnamefont {V.}~\bibnamefont {Mik{\v s}i{\'c}~Trontl}}, \bibinfo {author} {\bibfnamefont {I.}~\bibnamefont {Avigo}}, \bibinfo {author} {\bibfnamefont {P.}~\bibnamefont {Zhou}}, \bibinfo {author} {\bibfnamefont {M.}~\bibnamefont {Ligges}}, \bibinfo {author} {\bibfnamefont {D.}~\bibnamefont {{von Der Linde}}}, \bibinfo {author} {\bibfnamefont {U.}~\bibnamefont {Bovensiepen}}, \bibinfo {author} {\bibfnamefont {M.}~\bibnamefont {{Horn-von Hoegen}}}, \bibinfo {author} {\bibfnamefont {S.}~\bibnamefont {Wippermann}}, \bibinfo {author} {\bibfnamefont {A.}~\bibnamefont {L{\"u}cke}}, \bibinfo
  {author} {\bibfnamefont {S.}~\bibnamefont {Sanna}}, \bibinfo {author} {\bibfnamefont {U.}~\bibnamefont {Gerstmann}}, \ and\ \bibinfo {author} {\bibfnamefont {W.~G.}\ \bibnamefont {Schmidt}},\ }\bibfield  {title} {\enquote {\bibinfo {title} {Optically excited structural transition in atomic wires on surfaces at the quantum limit},}\ }\href {\doibase 10.1038/nature21432} {\bibfield  {journal} {\bibinfo  {journal} {Nature}\ }\textbf {\bibinfo {volume} {544}},\ \bibinfo {pages} {207--211} (\bibinfo {year} {2017})}\BibitemShut {NoStop}%
\bibitem [{\citenamefont {Hafke}\ \emph {et~al.}(2019)\citenamefont {Hafke}, \citenamefont {Witte}, \citenamefont {Brand}, \citenamefont {Duden},\ and\ \citenamefont {{Horn-von Hoegen}}}]{hafke_pulsed_2019}%
  \BibitemOpen
  \bibfield  {author} {\bibinfo {author} {\bibfnamefont {B.}~\bibnamefont {Hafke}}, \bibinfo {author} {\bibfnamefont {T.}~\bibnamefont {Witte}}, \bibinfo {author} {\bibfnamefont {C.}~\bibnamefont {Brand}}, \bibinfo {author} {\bibfnamefont {{\relax Th}.}~\bibnamefont {Duden}}, \ and\ \bibinfo {author} {\bibfnamefont {M.}~\bibnamefont {{Horn-von Hoegen}}},\ }\bibfield  {title} {\enquote {\bibinfo {title} {Pulsed electron gun for electron diffraction at surfaces with femtosecond temporal resolution and high coherence length},}\ }\href {\doibase 10.1063/1.5086124} {\bibfield  {journal} {\bibinfo  {journal} {Review of Scientific Instruments}\ }\textbf {\bibinfo {volume} {90}},\ \bibinfo {pages} {045119} (\bibinfo {year} {2019})}\BibitemShut {NoStop}%
\bibitem [{\citenamefont {{Hanisch-Blicharski}}\ \emph {et~al.}(2021)\citenamefont {{Hanisch-Blicharski}}, \citenamefont {Tinnemann}, \citenamefont {Wall}, \citenamefont {Thiemann}, \citenamefont {Groven}, \citenamefont {Fortmann}, \citenamefont {Tajik}, \citenamefont {Brand}, \citenamefont {Frost}, \citenamefont {{von Hoegen}},\ and\ \citenamefont {{Horn-von Hoegen}}}]{hanisch-blicharski_violation_2021}%
  \BibitemOpen
  \bibfield  {author} {\bibinfo {author} {\bibfnamefont {A.}~\bibnamefont {{Hanisch-Blicharski}}}, \bibinfo {author} {\bibfnamefont {V.}~\bibnamefont {Tinnemann}}, \bibinfo {author} {\bibfnamefont {S.}~\bibnamefont {Wall}}, \bibinfo {author} {\bibfnamefont {F.}~\bibnamefont {Thiemann}}, \bibinfo {author} {\bibfnamefont {T.}~\bibnamefont {Groven}}, \bibinfo {author} {\bibfnamefont {J.}~\bibnamefont {Fortmann}}, \bibinfo {author} {\bibfnamefont {M.}~\bibnamefont {Tajik}}, \bibinfo {author} {\bibfnamefont {C.}~\bibnamefont {Brand}}, \bibinfo {author} {\bibfnamefont {B.-O.}\ \bibnamefont {Frost}}, \bibinfo {author} {\bibfnamefont {A.}~\bibnamefont {{von Hoegen}}}, \ and\ \bibinfo {author} {\bibfnamefont {M.}~\bibnamefont {{Horn-von Hoegen}}},\ }\bibfield  {title} {\enquote {\bibinfo {title} {Violation of {{Boltzmann Equipartition Theorem}} in {{Angular Phonon Phase Space Slows}} down {{Nanoscale Heat Transfer}} in {{Ultrathin Heterofilms}}},}\ }\href {\doibase 10.1021/acs.nanolett.1c01665} {\bibfield  {journal}
  {\bibinfo  {journal} {Nano Letters}\ }\textbf {\bibinfo {volume} {21}},\ \bibinfo {pages} {7145--7151} (\bibinfo {year} {2021})}\BibitemShut {NoStop}%
\bibitem [{\citenamefont {{Horn-von Hoegen}}(2024)}]{horn-vonhoegenStructuralDynamicsSurfaces2024}%
  \BibitemOpen
  \bibfield  {author} {\bibinfo {author} {\bibfnamefont {M.}~\bibnamefont {{Horn-von Hoegen}}},\ }\bibfield  {title} {\enquote {\bibinfo {title} {Structural dynamics at surfaces by ultrafast reflection high-energy electron diffraction},}\ }\href {\doibase 10.1063/4.0000234} {\bibfield  {journal} {\bibinfo  {journal} {Structural Dynamics}\ }\textbf {\bibinfo {volume} {11}},\ \bibinfo {pages} {021301} (\bibinfo {year} {2024})}\BibitemShut {NoStop}%
\bibitem [{\citenamefont {Tietz}\ \emph {et~al.}(2022)\citenamefont {Tietz}, \citenamefont {Oster}, \citenamefont {Wisnet},\ and\ \citenamefont {Tietz}}]{tietzAdvantagesFiberOpticallyCMOS2022}%
  \BibitemOpen
  \bibfield  {author} {\bibinfo {author} {\bibfnamefont {H.}~\bibnamefont {Tietz}}, \bibinfo {author} {\bibfnamefont {M.}~\bibnamefont {Oster}}, \bibinfo {author} {\bibfnamefont {A.}~\bibnamefont {Wisnet}}, \ and\ \bibinfo {author} {\bibfnamefont {D.}~\bibnamefont {Tietz}},\ }\href@noop {} {\enquote {\bibinfo {title} {Advantages of {{Fiber-Optically CMOS Cameras}} for {{LEEM}}/{{PEEM Applications}}},}\ } (\bibinfo {year} {2022})\BibitemShut {NoStop}%
\bibitem [{\citenamefont {Shi}\ \emph {et~al.}(2016)\citenamefont {Shi}, \citenamefont {Nannenga}, \citenamefont {De~La~Cruz}, \citenamefont {Liu}, \citenamefont {Sawtelle}, \citenamefont {Calero}, \citenamefont {Reyes}, \citenamefont {Hattne},\ and\ \citenamefont {Gonen}}]{shiCollectionMicroEDData2016}%
  \BibitemOpen
  \bibfield  {author} {\bibinfo {author} {\bibfnamefont {D.}~\bibnamefont {Shi}}, \bibinfo {author} {\bibfnamefont {B.~L.}\ \bibnamefont {Nannenga}}, \bibinfo {author} {\bibfnamefont {M.~J.}\ \bibnamefont {De~La~Cruz}}, \bibinfo {author} {\bibfnamefont {J.}~\bibnamefont {Liu}}, \bibinfo {author} {\bibfnamefont {S.}~\bibnamefont {Sawtelle}}, \bibinfo {author} {\bibfnamefont {G.}~\bibnamefont {Calero}}, \bibinfo {author} {\bibfnamefont {F.~E.}\ \bibnamefont {Reyes}}, \bibinfo {author} {\bibfnamefont {J.}~\bibnamefont {Hattne}}, \ and\ \bibinfo {author} {\bibfnamefont {T.}~\bibnamefont {Gonen}},\ }\bibfield  {title} {\enquote {\bibinfo {title} {The collection of {{MicroED}} data for macromolecular crystallography},}\ }\href {\doibase 10.1038/nprot.2016.046} {\bibfield  {journal} {\bibinfo  {journal} {Nature Protocols}\ }\textbf {\bibinfo {volume} {11}},\ \bibinfo {pages} {895--904} (\bibinfo {year} {2016})}\BibitemShut {NoStop}%
\bibitem [{\citenamefont {Inc}(2001)}]{burleindustriesincProceduresSpecicationsAPD2001}%
  \BibitemOpen
  \bibfield  {author} {\bibinfo {author} {\bibfnamefont {B.~I.}\ \bibnamefont {Inc}},\ }\href@noop {} {\enquote {\bibinfo {title} {Procedures and {{Specications}}: {{APD 3040FM}} 12/10/8 l 60:1 {{EDR P20 6FL}}. (2001)},}\ } (\bibinfo {year} {2001})\BibitemShut {NoStop}%
\bibitem [{\citenamefont {Woodhead}\ and\ \citenamefont {Ward}(1977)}]{woodheadChannelElectronMultiplier1977}%
  \BibitemOpen
  \bibfield  {author} {\bibinfo {author} {\bibfnamefont {A.}~\bibnamefont {Woodhead}}\ and\ \bibinfo {author} {\bibfnamefont {R.}~\bibnamefont {Ward}},\ }\bibfield  {title} {\enquote {\bibinfo {title} {The channel electron multiplier and its use in image intensifiers},}\ }\href {\doibase 10.1049/ree.1977.0079} {\bibfield  {journal} {\bibinfo  {journal} {Radio and Electronic Engineer}\ }\textbf {\bibinfo {volume} {47}},\ \bibinfo {pages} {545--553} (\bibinfo {year} {1977})}\BibitemShut {NoStop}%
\bibitem [{\citenamefont {Klingelhoefer}, \citenamefont {Wiacker},\ and\ \citenamefont {Kankeleit}(1986)}]{klingelhoeferMeasurementDetectionEfficiency1986}%
  \BibitemOpen
  \bibfield  {author} {\bibinfo {author} {\bibfnamefont {G.}~\bibnamefont {Klingelhoefer}}, \bibinfo {author} {\bibfnamefont {H.}~\bibnamefont {Wiacker}}, \ and\ \bibinfo {author} {\bibfnamefont {E.}~\bibnamefont {Kankeleit}},\ }\bibfield  {title} {\enquote {\bibinfo {title} {Measurement of the detection efficiency of microchannel plates for 1--15 {{keV}} electrons},}\ }\href {\doibase 10.1016/0168-9002(86)91320-3} {\bibfield  {journal} {\bibinfo  {journal} {Nuclear Instruments and Methods in Physics Research Section A: Accelerators, Spectrometers, Detectors and Associated Equipment}\ }\textbf {\bibinfo {volume} {247}},\ \bibinfo {pages} {379--384} (\bibinfo {year} {1986})}\BibitemShut {NoStop}%
\bibitem [{\citenamefont {Richter}\ and\ \citenamefont {Ho}(1986)}]{richterPositionsensitiveDetectorPerformance1986}%
  \BibitemOpen
  \bibfield  {author} {\bibinfo {author} {\bibfnamefont {L.~J.}\ \bibnamefont {Richter}}\ and\ \bibinfo {author} {\bibfnamefont {W.}~\bibnamefont {Ho}},\ }\bibfield  {title} {\enquote {\bibinfo {title} {Position-sensitive detector performance and relevance to time-resolved electron energy loss spectroscopy},}\ }\href {\doibase 10.1063/1.1138572} {\bibfield  {journal} {\bibinfo  {journal} {Review of Scientific Instruments}\ }\textbf {\bibinfo {volume} {57}},\ \bibinfo {pages} {1469--1482} (\bibinfo {year} {1986})}\BibitemShut {NoStop}%
\bibitem [{\citenamefont {Henzler}(1984)}]{henzlerMeasurementSurfaceDefects1984a}%
  \BibitemOpen
  \bibfield  {author} {\bibinfo {author} {\bibfnamefont {M.}~\bibnamefont {Henzler}},\ }\bibfield  {title} {\enquote {\bibinfo {title} {Measurement of surface defects by low-energy electron diffraction},}\ }\href {\doibase 10.1007/BF00616574} {\bibfield  {journal} {\bibinfo  {journal} {Applied Physics A}\ }\textbf {\bibinfo {volume} {34}},\ \bibinfo {pages} {205--214} (\bibinfo {year} {1984})}\BibitemShut {NoStop}%
\bibitem [{\citenamefont {{Horn-von Hoegen}}(1999)}]{horn-vonhoegenGrowthSemiconductorLayers1999}%
  \BibitemOpen
  \bibfield  {author} {\bibinfo {author} {\bibfnamefont {M.}~\bibnamefont {{Horn-von Hoegen}}},\ }\bibfield  {title} {\enquote {\bibinfo {title} {Growth of semiconductor layers studied by spot profile analysing low energy electron diffraction -- {{Part I1}}},}\ }\href {\doibase 10.1524/zkri.1999.214.10.591} {\bibfield  {journal} {\bibinfo  {journal} {Zeitschrift f{\"u}r Kristallographie - Crystalline Materials}\ }\textbf {\bibinfo {volume} {214}},\ \bibinfo {pages} {591--629} (\bibinfo {year} {1999})}\BibitemShut {NoStop}%
\bibitem [{\citenamefont {Klein}\ \emph {et~al.}(2011)\citenamefont {Klein}, \citenamefont {Nabbefeld}, \citenamefont {Hattab}, \citenamefont {Meyer}, \citenamefont {Jnawali}, \citenamefont {Kammler}, \citenamefont {{zu Heringdorf}}, \citenamefont {{Golla-Franz}}, \citenamefont {M{\"u}ller}, \citenamefont {Schmidt}, \citenamefont {Henzler},\ and\ \citenamefont {{Horn-von Hoegen}}}]{kleinLostReciprocalSpace2011}%
  \BibitemOpen
  \bibfield  {author} {\bibinfo {author} {\bibfnamefont {C.}~\bibnamefont {Klein}}, \bibinfo {author} {\bibfnamefont {T.}~\bibnamefont {Nabbefeld}}, \bibinfo {author} {\bibfnamefont {H.}~\bibnamefont {Hattab}}, \bibinfo {author} {\bibfnamefont {D.}~\bibnamefont {Meyer}}, \bibinfo {author} {\bibfnamefont {G.}~\bibnamefont {Jnawali}}, \bibinfo {author} {\bibfnamefont {M.}~\bibnamefont {Kammler}}, \bibinfo {author} {\bibfnamefont {F.-J.~M.}\ \bibnamefont {{zu Heringdorf}}}, \bibinfo {author} {\bibfnamefont {A.}~\bibnamefont {{Golla-Franz}}}, \bibinfo {author} {\bibfnamefont {B.~H.}\ \bibnamefont {M{\"u}ller}}, \bibinfo {author} {\bibfnamefont {{\relax Th}.}~\bibnamefont {Schmidt}}, \bibinfo {author} {\bibfnamefont {M.}~\bibnamefont {Henzler}}, \ and\ \bibinfo {author} {\bibfnamefont {M.}~\bibnamefont {{Horn-von Hoegen}}},\ }\bibfield  {title} {\enquote {\bibinfo {title} {Lost in reciprocal space? {{Determination}} of the scattering condition in spot profile analysis low-energy electron diffraction},}\ }\href
  {\doibase 10.1063/1.3554305} {\bibfield  {journal} {\bibinfo  {journal} {Review of Scientific Instruments}\ }\textbf {\bibinfo {volume} {82}},\ \bibinfo {pages} {035111} (\bibinfo {year} {2011})}\BibitemShut {NoStop}%
\bibitem [{App(1999)}]{AppliedRHEED1999}%
  \BibitemOpen
  \href {\doibase 10.1007/BFb0109548} {\emph {\bibinfo {title} {Applied {{RHEED}}}}},\ \bibinfo {series} {Springer {{Tracts}} in {{Modern Physics}}}, Vol.\ \bibinfo {volume} {154}\ (\bibinfo  {publisher} {Springer},\ \bibinfo {address} {Berlin, Heidelberg},\ \bibinfo {year} {1999})\BibitemShut {NoStop}%
\end{thebibliography}%

Copyright (2024) Jonas D. Fortmann, Alexander Kaßen, Christian Brand, Thomas Duden \& Michael Horn-von Hoegen. This article is distributed under a Creative Commons Attribution (CC BY-NC-ND) License

\end{document}